\documentclass[final]{aipproc}
\layoutstyle{6x9}

\newcommand{\ave}[1]{{\langle #1\rangle}}

\begin{document}

\title[Control of Heat Flow]{Heat Flow in Classical and Quantum Systems and Thermal Rectification}


\author{Giulio Casati}{address={Center for  Nonlinear and  Complex  Systems,
  Universit\`a degli Studi dell'Insubria, Como, Italy, CNR-INFM; Istituto
  Nazionale di Fisica Nucleare, Sezione di Milano and Department of Physics;
  National University of Singapore, Singapore 117542, Republic of Singapore}}
\author{Carlos Mej\'{\i}a-Monasterio}{address={Dipartimento di Matematica, Politecnico  di Torino, Torino, Italy}}

\begin{abstract}
  The understanding of the underlying dynamical mechanisms which determine the
  macroscopic  laws   of  heat   conduction  is  a   long  standing   task  of
  non-equilibrium  statistical  mechanics.   A  better  understanding  of  the
  mechanism   of  heat   conduction  may   lead  to   potentially  interesting
  applications  based on  the possibility  to control  the heat  flow. Indeed,
  different models of  thermal rectifiers has been recently  proposed in which
  heat can  flow preferentially in  one direction.  Although these  models are
  far away from a prototype realization, the underlying mechanisms are of very
  general nature and, as such,  are suitable of improvement and may eventually
  lead to real applications. We  briefly discuss the problem of heat transport
  in classical and  quantum systems and its relation to  the chaoticity of the
  dynamics. We then study the  phenomenon of thermal rectification and briefly
  discuss  the different  types of  microscopic  mechanisms that  lead to  the
  rectification of heat flow.
\end{abstract}

\maketitle

\section{Introduction}
\label{sec:intro}

The origin of the macroscopic  phenomenological laws of transport is still one
of the major  challenges in theoretical physics.  In  particular, the issue of
heat transport, in  spite of having a long history,  is not completely settled
\cite{bonetto,LLP-review}.    Given    a   particular   classical,   many-body
Hamiltonian system, neither  phenomenological nor fundamental transport theory
can predict whether  or not this specific Hamiltonian  system yields an energy
transport  governed  by the  Fourier  law  $J=-\kappa\nabla  T$, relating  the
macroscopic heat flux to the temperature gradient $\nabla T$ \cite{peierls}.

In     spite     of     intense     investigations    in     recent
years,
\cite{fl1,leprifpu,hatano99,Dhar01,alonso,triangle1,casati,mejia-1,mejia-2,emmz}
the precise conditions that a  dynamical system of interacting
particles must satisfy in order to obey the Fourier law of heat
conduction are still not known. However, the general picture that
emerges is that, for systems with no globally  conserved  quantities
(i.e.,  globally  ergodic), positive  Lyapunov exponents  is a
sufficient  condition to  ensure the  Fourier heat  law.

Given the  state of affairs,  a common strategy  is to investigate
up  to what extent  one can  simplify the  microscopic dynamics  and
yet  obtain  a normal transport behavior.  Thermal transport has
been studied for a Lorentz channel --a quasi-one-dimensional
billiard with circular scatterers-- and it was shown to obey the
Fourier law \cite{alonso}, yet we do not have rigorous results and
in spite of several efforts,  the connection between Lyapunov
exponents, decay of correlations and diffusive and transport
properties is still not completely clear.  For example, in a recent
paper \cite{triangle1}, a model was presented which  has  zero
Lyapunov  exponents,  yet it  exhibits  unbounded  Gaussian
diffusive behavior.   Since diffusive behavior is  at the root  of
normal heat transport, the above  result constitutes a strong
suggestion  that normal heat conduction can take  place even without
the strong requirement of exponential instability.   The  models in
Refs.~\cite{alonso} and  \cite{triangle1}  are noninteracting: thus,
the condition of  Local Thermal Equilibrium (LTE) is not satisfied.

At the quantum level, the question whether normal transport may
arise from the underlying quantum dynamics remains an open issue
\cite{qfl-list,qfl}. This is mostly because it  is not clear how to
describe the  transport of energy or heat from  a microscopic point
of view.  In  analogy to classical  systems, a quantum derivation of
the Fourier law  calls directly in question the issue of quantum
chaos  \cite{rmt}.  However, a main  feature of quantum  motion is
the lack  of exponential  dynamical  instability \cite{casati86}.
This fact  may render very  questionable the  possibility to derive
the Fourier law  of heat conduction in  quantum mechanics.  Thus it
is interesting to  inquire if, and under what conditions, Fourier
law  emerges from the laws of quantum mechanics (for a recent review
of the microscopic foundations of the quantum Fourier law see
\cite{qfl-rev}).

In this  paper we present  a brief review  of heat transport in  classical and
quantum systems.   We then investigate  the possibility to control  the energy
transport and discuss different microscopic mechanical models in which thermal
rectification can be observed.  The possibility of controlling heat conduction
by nonlinearity  opens the way  to design a  thermal rectifier, {\em  i.e.}, a
system that carries  heat preferentially in one direction  while it behaves as
an insulator in the opposite direction.

\section{Dynamical Instability and Fourier Law}
\label{sec:fourier}

\subsection{Fourier Law in Classical Systems}
\label{sec:classical}

To clarify the  role of dynamical instability for the  validity of
Fourier law, different  microscopic mechanical  models  ({\em e.g.},
chains of  anharmonic oscillators and billiards)  have been
investigated. Here we focus on billiards with integrable or chaotic
dynamics.

\begin{figure}[!t]
\includegraphics[width=5.5in]{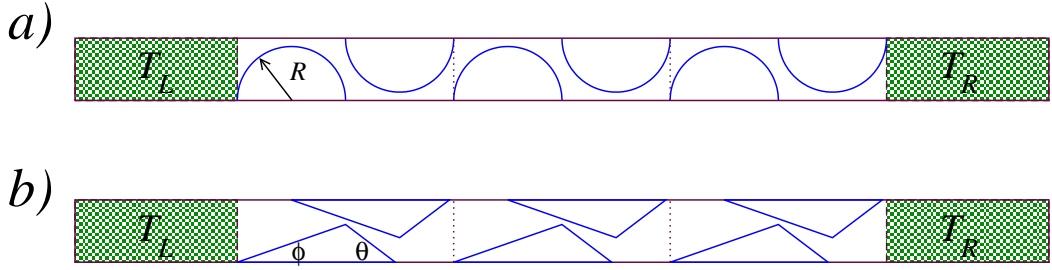}
\caption {
  Geometry  of different  billiards: $a$)  a Lorentz  gas channel  at critical
  horizon,  {\em i.e.},  the  separation among  the  centers of  the discs  is
  $4R/\sqrt{3}$,  and $b$) a triangle billiard  channel whose geometry
  is uniquely specified by assigning the internal angles $\theta$ and $\phi$.
\label{fig:billiards}}
\end{figure}

Motivated  by the  ergodicity and  mixing properties  of the Lorentz
gas, in Ref.~\cite{alonso} a  channel geometry has been considered
to study the problem of heat conduction in this model. A Lorentz gas
consists of noninteracting point particles that  collide elastically
with a set of circular  scatterers on the plane  (see
Fig.~\ref{fig:billiards}-$a$).   By imposing  an  external thermal
gradient,  it  was  found that  in  a  Lorentz  channel  the Fourier
law  was satisfied, yet  this system is not  described by LTE.  A
modification of this model  in which  particles and  scatterers can
exchange energy  through their collisions appeared in
Ref.~\cite{mejia-1}.  This effective interaction leads to  the
establishment  of LTE.   As a  consequence, this  model has  proven
to reproduce  realistically macroscopic  transport in  many
different situations \cite{mejia-2}.  In particular, the validity of
the Fourier law was verified. One  can  conclude  that,  while the
interaction  among  particles  is  not fundamental for the
observation of  normal transport, it is strictly necessary for  the
identification  of   microscopic  dynamical  quantities   with  the
macroscopic physical parameters.

More  recently, in order to  investigate  if chaos  is  a necessary
condition for  the validity of  Fourier law,  a two-dimensional
billiard  model - which  consists of a rectangular area  and a
series of triangular  scatterers- was  considered (see
Fig.~\ref{fig:billiards}-$b$).   This model  is analogous  to the
Lorentz gas channel with triangles instead of  discs, and the
essential difference is that in  the  triangle  billiard  channel,
the dynamical  instability  is  linear; therefore the Lyapunov
exponent is zero.  Strong numerical  evidence has been recently
given \cite{casati} that the  motion inside a triangle billiard,
with all angles irrational with $\pi$, is mixing, without any time
scale, (see also \cite{poly}).   Moreover, an  area-preserving  map,
which  was  derived as  an approximation of the boundary map for the
irrational triangle, shows a Gaussian diffusive behavior when
considered on the cylinder, even though the Lyapunov exponent of the
map is zero \cite{triangle1}. It is therefore reasonable to expect
that the  motion inside  the irrational  triangle billiard channel
of Fig.~\ref{fig:billiards}-$b$   is   diffusive,    thus leading to
normal conductivity.

We have performed numerical simulations ~\cite{triang} on a triangle
billiard channel of total length $L =Nl$, where $N$ and $l$ are  the
number and the length of the fundamental cells, as indicated in
Fig~\ref{fig:billiards}.   Heat  baths have been simulated with
stochastic kernels of Gaussian type: namely, the probability
distribution of velocities for particles coming out from the baths
is
\begin{equation} \label{eq:baths}
P(v_x) = \frac{|v_x|}{T}\exp\left(-\frac{v^2_x}{2T}\right), \
P(v_y) = \frac{1}{\sqrt{2\pi T}}\exp\left(-\frac{v^2_y}{2T}\right)
\label{Gaussian}
\end{equation}
for  $v_x$  and  $v_y$,  respectively.   Since  the  energy  changes  only  at
collisions with the heat baths, the heat flux is given by
\begin{equation}
j (t_c)=\frac{1}{t_c}\sum_{k=1}^{N_c}(\Delta E)_k , \label{flux}
\end{equation}
where $(\Delta E)_k =  E_{in} - E_{out}$ is the change in  energy at the $k$th
collision with the heat bath and  $N_c$ is the total number of such collisions
that  occur during time  $t_c$.

\begin{figure}[!t]
\includegraphics[scale=0.75]{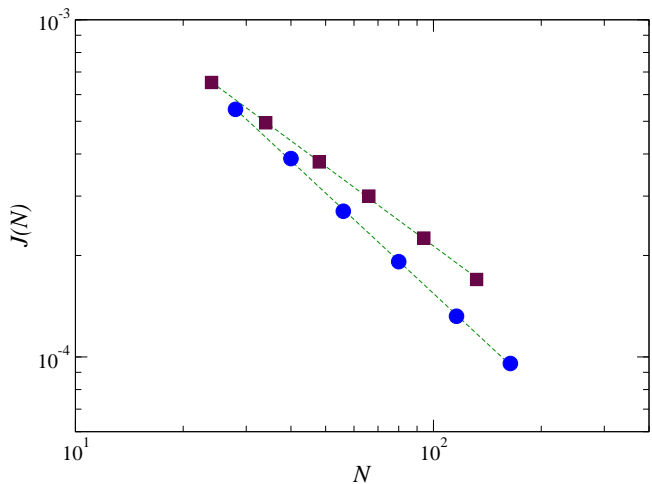}
\caption{
  Scaling behavior of the stationary heat flux $J$ as a function of the system
  size  for the case  of irrational  angles of  Fig. 1  (circles) and  for the
  rational  case (squares).   $N$ is  the number  of fundamental  cells. The
  particle density was set to $1$ particle per cell independently of $N$.  The
  least-squares fitting  gives a  slope $ -0.99  \pm 0.01$ for  the irrational
  case and $-0.78\pm0.01$ for the rational one.}
\label{heatflux}
\end{figure}

In Fig.~\ref{heatflux}, the heat flux $J$ as a function of the
system size $N$ is  shown.  For  the  case of  irrational angles
($\theta=(\sqrt{2}-1)\pi/2$, $\phi=1$), the best  fit gives
$J\propto N^{-\gamma}$, with  $\gamma = 0.99\pm 0.01$. The
coefficient of thermal conductivity is,  therefore, independent on
$N$, which means  that the Fourier law is obeyed. The  same result
is obtained using the Green-Kubo formalism \cite{triang}.  A
completely different behavior is  obtained when the  angles $\theta$
and $\phi$  are rational  multiples of $\pi$.    The  case  with
$\theta=\pi/5$  and   $\phi=\pi/3$  is   shown  in
Fig.~\ref{heatflux}  (triangles),  from  which  a divergent behavior
of  the coefficient  of thermal conductivity,  $ \kappa \sim
N^{0.22}$,  is observed, indicating the absence of the Fourier law.

In  conclusion,  when  all  angles  are irrational  multiples  of  $\pi$,  the
triangle  billiard  channel exhibits  the  Fourier  law  of heat  conduction
together  with  nice  diffusive  properties.   However, when  all  angles  are
rational multiple  of $\pi$, the model  shows abnormal diffusion  and the heat
conduction does not follow the Fourier law.

One may  argue that the  model considered here  is somehow
artificial  and far from a realistic physical  model.  Indeed,
noninteracting particles systems are certainly less realistic  as,
in general, LTE is not  established. In order to elucidate the  role
of  interactions, we have  studied in \cite{caspr}  and in
\cite{caspb}  the  so-called \emph{1d  hard-point  particles with
alternating
  masses}  with and  without total  momentum conservation  respectively.  This
model,  consists of  a  one-dimensional chain  of  elastically colliding  free
particles with alternate masses  $m$ and $M$ \cite{hatano99,Dhar01}. It shares
with the triangle billiard the linear dynamical instability but in contrast,
it is a genuinely interacting many-particle system.

When  the total  momentum is  conserved we  have found  that the  heat current
scales as $J\propto N^{-\alpha}$  with $\alpha\sim0.745$ and thus, in contrast
with  the  irrational  triangle  channel,  the  alternating  mass  model  with
conservation of total momentum does not obey the Fourier law \cite{caspr}.  We
recall that  in several recent papers  \cite{hatano99,Campbell,narayan} it has
been suggested that total momentum conservation does not allow Fourier law and
this may explain  the lack of Fourier law for  the one dimensional alternating
mass  model. Indeed, in  \cite{caspb} we  have found  that the  alternate mass
hard-point gas without total momentum conservation obeys the Fourier law.

In perspective, these results  demonstrate that diffusive energy
transport and Fourier law can take  place in marginally stable
(non-chaotic) interacting many-particles  systems.    As  a
consequence,   the  exponential  instability (Lyapunov chaos)  is
not necessary for  the establishment of the  Fourier law.
Furthermore, our results show that breaking  total momentum
conservation is crucial for the validity of Fourier law.

\subsection{Fourier Law in Quantum Systems}
\label{sec:quantum}

In the  previous section, we  have shown that strong,  exponentially
unstable, classical chaos  is not  necessary (actually, strictly
speaking, is  not even sufficient \cite{leprifpu})  for normal
transport. In this  connection we remark that a main feature of
quantum motion is the lack  of exponential  dynamical instability
\cite{casati}. Thus,  it  is  interesting  to inquire  if,  and
under  what conditions, the Fourier law emerges from the laws of
quantum mechanics.

\begin{figure}[!t]
\includegraphics[scale=0.6]{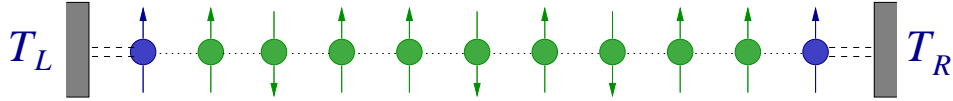}
\caption{Schematic representation of a finite one-dimensional quantum spin
  chain, coupled to external heat reservoirs at different temperatures.
\label{fig:chain}}
\end{figure}

To this end we consider  an Ising chain  of $L$ spins  $1/2$ with a
coupling constant $Q$ subject  to  a uniform  magnetic  field
$\vec{h}  = (h_x,0,h_z)$,  with  open boundaries. The Hamiltonian
reads
\begin{equation} \label{eq:H}
{\mathcal H} = -Q\sum_{n=0}^{L-2}\sigma^z_n\sigma^z_{n+1} +
\vec{h}\cdot\sum_{n=0}^{L-1}\vec{\sigma}_n \ ,
\end{equation}
where the operators  $\vec{\sigma}_n = (\sigma^x_n,\sigma^y_n,\sigma^z_n)$ are
the  Pauli  matrices for  the  $n$th spin,  $n=0,1,\ldots  L-1$.   We set  the
coupling constant $Q=2$. A schematic  representation of this model is shown in
Fig.~\ref{fig:chain}.   In  this  system,  the  only  trivial  symmetry  is  a
reflection    symmetry,    $\vec{\sigma}_n\rightarrow\vec{\sigma}_{L-1-n}$.
Moreover, the direction of the magnetic field affects the qualitative behavior
of the system: If $h_z=0$, the Hamiltonian Eq.~\eqref{eq:H} corresponds to the
Ising chain  in a  transversal magnetic  field.  In this  case, the  system is
integrable as it can be mapped  into a model of free fermions through standard
Wigner-Jordan transformations.  When $h_z$  is increased from zero, the system
is no longer integrable, and when $h_z$ is of the same order of $h_x$, quantum
chaos sets in,  leading to a very complex structure of  quantum states as well
as to fluctuations in the  spectrum that are statistically described by Random
Matrix Theory (RMT) \cite{rmt}.   The system becomes again (nearly) integrable
when  $h_z \gg h_x$.   Therefore, by  choosing the  direction of  the external
field, we can explore different regimes of quantum dynamics.

The transition to quantum chaos for this model has been studied in \cite{qfl}.
We   consider    three   cases:   ({\it   i})   the    {\em   chaotic   chain}
$\vec{h}=(3.375,0,2)$,    ({\it    ii})    the    {\em    integrable    chain}
$\vec{h}=(3.375,0,0)$,   and  ({\it   iii})  the   {\em   intermediate  chain}
$\vec{h}=(7.875,0,2)$ which is neither chaotic nor integrable.

In  Ref.~\cite{qfl} we  have simulated  the coupling  of the  spin  chain with
thermal baths, requiring  that the state of the spin in  contact with the bath
is statistically  determined by a  Boltzmann distribution with parameter  $T$.
Our model for the reservoirs is analogous to the stochastic thermal reservoirs
defined by  Eqs.~\eqref{eq:baths}, thus, we  call it a  \emph{quantum stochastic
  reservoir}.   In what  follows we  use  units in  which the  Planck and  the
Boltzmann constants are set to unity, $\hbar=k_{\rm B}=1$.

The  dynamics of the  spins is  obtained from  the unitary  evolution operator
$\mathrm{U}(t)  = \exp(-i\mathcal{H}t)$.  Additionally,  the leftmost  and the
rightmost spins of  the chain are coupled to  quantum stochastic reservoirs at
temperatures $\beta_T^{-1}$ and  $\beta_R^{-1}$, respectively. For the details
of  the   quantum  stochastic   reservoir  model  we   refer  the   reader  to
Ref.~\cite{qfl} and for  a comparison with the solution  of the quantum master
equation in Lindblad form to \cite{MMW}.

In  order  to  compute  the  energy  profile,  we  write  the  Hamiltonian  in
Eq.~\eqref{eq:H} as
\begin{equation} \label{eq:Htot}
\mathcal{H} = \sum_{n=0}^{L-2}H_n +
\frac{h}{2}(\sigma_\lambda + \sigma_\rho) \ ,
\end{equation}
where   $\sigma_\lambda   =   \vec{h}\cdot\vec{\sigma}_0/h$,  $\sigma_\rho =  \vec{h}\cdot\vec{\sigma}_{L-1}/h$ are boundary terms and
\begin{equation} \label{eq:H_local}
H_{n}     =   -Q\sigma^z_n\sigma^z_{n+1}    +
\frac{\vec{h}}{2} \cdot \left(\vec{\sigma}_n  + \vec{\sigma}_{n+1}\right) \ ,
\quad 0 <  n <  L-2 \ ,
\end{equation}
are the local energy density operators for the $n$th and $(n+1)$th
spins.

The local  current operators are defined through the equation of
continuity: $\partial_t{H_n} =  i[{\mathcal H},H_n] =  - (J_{n+1} -
J_n)$,  requiring that $J_n = [H_n,H_{n-1}]$.  With
Eqs.~\eqref{eq:H_local} and \eqref{eq:Htot} the local current
operators are explicitly given by
\begin{equation}
J_{n} = h_xQ\left(\sigma_{n-1}^z-\sigma_{n+1}^z\right)\sigma^y_{n},
\quad
1\le n\le  L-2.
\end{equation}

\begin{figure}[!t]
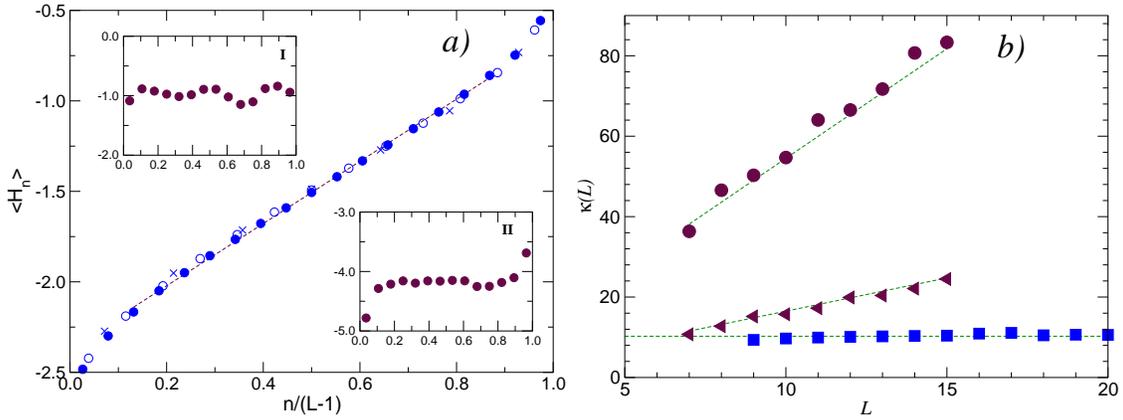

\includegraphics[scale=0.65]{qfl-T.eps}
\includegraphics[scale=0.65]{qfl.eps}
\caption{
  $a$) Energy profile $\ave{H_n}$ for  the chaotic chain.  The temperatures of
  the baths were set to $T_L = 5$  and $T_R = 50$.  Results for chains of size
  $L=8$ (crosses), $L=14$ (open circles) and $L=20$ (solid circles) are shown.
  The dashed line  was obtained from a  linear fit of the data  for $L=20$ for
  the $L-4$  central spins. Insets (I)  and (II) show the  energy profiles for
  the integrable  and the intermediate  cases, respectively, for  $L=15$. $b$)
  Heat  conductivity  as  a  function  of  the system  size  for  the  chaotic
  (squares), integrable (triangles) and intermediate (circles) chains.
\label{fig:3}}
\end{figure}

In  Fig.~\ref{fig:3}-$a$,   the  energy  profile   for  an  out-of-equilibrium
simulation of the  chaotic chain is shown.  After  an appropriate scaling, the
profiles for  different sizes collapse to the  same curve which is,  to a very
good approximation, linear.  In contrast, in the case of the integrable (inset
I) and the  intermediate (inset II) chains, no energy  gradient is created. In
Fig.~\ref{fig:3}-$b$, the heat conductivity  $\kappa = J/\nabla(1/T)$ is shown
as a function of the size of the chain. The constant value of $\kappa$ for the
chaotic spin  chain indicates that the  Fourier law is satisfied,  while it is
violated for the integrable and intermediate chains.

In  conclusion  these results  suggest  that the  onset  of  quantum chaos  is
required for the validity of the Fourier law.

\section{Thermal Rectification}
\label{sec:rectifier}

We  now turn  our  attention to  the  possibility of  controlling
the heat flow. Consider a system subjected to an external
temperature gradient $\nabla T$, so that a  stationary and uniform
heat current $J^+$ appears,  transporting heat from the hot to the
cold reservoir. Then consider the case in which the temperature
gradient is inverted to $-\nabla T$ and  the heat current $J^-$  is
measured. In short, we say  that the system is a thermal rectifier
if $J^+ \ne J^-$. To quantify the power of rectification we use the
ratio of the two currents:
\begin{equation} \label{eq:rect}
\Delta = \frac{\max\{|J^+|,|J^-|\}}{\min\{|J^+|,|J^-|\}} ~.
\end{equation}

In  Ref.~\cite{diod} a  microscopic  mechanism for  thermal
rectification  was proposed  for the first  time. There,  an
anharmonically  interacting particles chain has been considered.
Using an effective phonon approach,  it was shown that by setting a
strongly nonlinear central region sandwiched  between two weakly
anharmonic left and  right  domains, the  anharmonic  chain exhibits
thermal rectification with ratio $\Delta\sim2$.  In \cite{diod} the
phenomenon was explained in terms of the (non)matching of the
effective phonon bands.

This and related ideas  were further elaborated and improved \cite{li-2,li-3},
achieving  rectification efficiencies up  to $\Delta\sim2000$.   Recently, the
ideas in  \cite{li-2} have led to  an interesting experimental  work in which,
thermal  rectification has been  observed \cite{science}.   A further  step to
devise  a thermal  transistor was  discussed in  \cite{li-4} in  terms  of the
negative differential thermal resistance observed in some anharmonic chains.

Other  different  mechanisms  leading   to  thermal  rectification
have  been described \cite{segal,EMM}. In particular, in \cite{EMM}
thermal rectification in asymmetric billiards  of interacting
particles was described  for the first time, and  rectifications as
large  as $\Delta\sim10^3$ were  observed. A simple phenomenological
mechanism  of thermal  rectification, has also  been recently
discussed \cite{phenom}.

The possibility  of obtaining  large $\Delta$ in  billiard systems  has raised
great interest because  they are more easily realizable  experimentally in the
rapidly expanding field of nanophysics. More recently we have proposed a novel
mechanism for thermal rectification  that results from the asymmetric behavior
of the dynamics at a magnetic interface \cite{magrec}.

\begin{figure}[!t]
\includegraphics[scale=0.55]{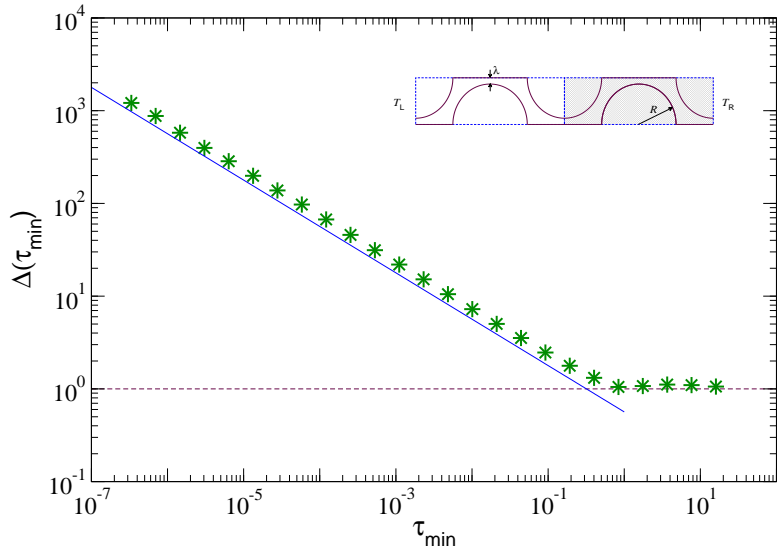}
\caption{
  Thermal  rectification $\Delta$  as a  function of  the  minimal temperature
  $\tau_{\min}$.  The maximal  temperature was  set to  $\tau_{\max}=33.4275$.
  The dashed  line is  for the no  rectification value $\Delta=1$.   The solid
  line corresponds  to $\tau_{\min}^{-1/2}$.  Inset:  schematic representation
  of the thermal rectifier.
\label{fig:magrec}}
\end{figure}

Consider  a gas of  noninteracting point  particles of  mass $m$ and
electric charge $e$ that  move freely inside a closed
two-dimensional billiard region. The   billiard   has   the geometry
of  the   Lorentz   gas   channel   of Fig.~\ref{fig:billiards}-$a$.
Furthermore,  we   break   the symmetry   by considering that  the
left  half of the  billiard contains no  magnetic field, whereas the
right cell is  subjected to a perpendicular uniform magnetic field
of strength  $B$. Finally, the  left and right  boundaries of the
billiard are coupled to stochastic heat baths (as in
Eq.~\eqref{eq:baths}), at temperatures $T_L$ and  $T_R$. A schematic
representation  of the billiard is shown in the inset of
Fig.~\ref{fig:magrec}.

It is worth mentioning that the appearance of rectification does not depend on
the particular geometry  of the billiard.  However, the  negative curvature of
the billiard boundary  ensures that the motion in the  absence of the magnetic
field is completely chaotic.

The  transmission probability  between  the  left and right cells
is controlled by  the strength of the magnetic field. Consider the
particles that cross the junction from  \emph{left to  right}. There
exist a critical  velocity $v_c$:  fast particles  of velocity  $v
> v_c$, always  enter the  right  cell, and  thus contribute to the
left to right  energy flow provided they reach the right end of the
system. Instead, slow particles of velocity $v < v_c$, such that the
gyro-magnetic  radius $\rho(v)  = mv/(eB)$  is less  than
$\lambda/2$  will be reflected or  transmitted depending on the
position  at which they  reach the interface.

Using a statistical  ensemble of trajectories, the condition  for the critical
velocity $\rho(v_c)  = \lambda/2$ can be  rewritten as the  condition giving a
critical temperature
\begin{equation} \label{eq:Tc}
T_{\mathrm{c}} = \frac{(eB_{\mathrm{c}} \lambda)^2}{8m k_{\rm
    B}} \ ,
\end{equation}
such that particles which are colder than $T_{\rm c}$ will be reflected in
their majority.

In contrast,  for the  particles that cross  the junction from  \emph{right to
  left} there is no condition on their velocity and they always enter the left
cell.  The  above qualitative  argument makes it  clear that the
transport of heat  will  be strongly  asymmetric  with  respect  to
exchange  of  effective temperatures  of particles  on  different
sides  of  magnetic field  boundary, provided the  temperatures are
strongly different, one  being larger  and the other  smaller than
$T_{\rm  c}$.  Denoting  with  $\tau =  T/T_{\rm c}$  the
temperature in units of $T_{\rm c}$,  it is then clear that
rectification will be effective if  one of the temperatures is very
small,  say $\tau_{\rm L} \ll 1$ and the other is simply above the
critical $\tau_{\rm R} > 1$.

We  measure the heat  current per  particle in  the steady  state as  the time
average of the energy transported across the junction per unit time
\begin{equation}
J = \lim_{t\rightarrow\infty} \frac{1}{t}\int_0^t
\frac{1}{N}\sum_{i=1}^N
E_i(s)\mathrm{sgn}(v_{x~i}(s))\delta(x_i-x_\mathrm{junc})\mathrm{d}s \ ,
\end{equation}
where $E_i(t)  = \frac{1}{2}mv_i^2(t)$ is the instantaneous  kinetic energy of
the $i$th  particle and $x_\mathrm{junc}$  the position along the  $x$-axis of
the junction. Furthermore,  we denote the heat current  as $J^+$ if $\tau_{\rm
  L} <  \tau_{\rm R}$ and as $J^{-}$  if the temperatures are  exchanged, i.e.
$\tau_{\rm L} > \tau_{\rm R}$.

In \cite{magrec}  it was  shown that,  if $\tau_{\rm L}  < \tau_{\rm  R}$, the
current $J^+$ is proportional to the transmission coefficient at the interface
$t^+$   and  invoking   ergodicity,  it   can  be   estimated  as   $t^+  \sim
\frac{2\rho(v)}{\lambda}$, where $\rho(v(\tau))=\sqrt{2mk_B\tau_{\rm min}}/eB$
and $\tau_{\rm min} = {\rm min}\{ \tau_{\rm L},\tau_{\rm R}\}$.

However, in  the reverse situation  (exchanging $\tau_{\rm L}$  and $\tau_{\rm
  R}$) we have $t^- \sim 1$, so the rectification becomes
\begin{equation} \label{eq:scaling}
\Delta = t^-/t^+ \propto \frac{1}{\sqrt{\tau_{\rm min}}} ~.
\end{equation}

In Fig.~\ref{fig:magrec} the rectification $\Delta$  is shown as a function of
$\tau_{\rm min}$ for fixed value  of the maximal temperature $\tau_{\rm max}$,
confirming  the estimate  of  Eq.~\eqref{eq:scaling}. The  correctness of  the
scaling \eqref{eq:scaling}  shows that the  magnetically induced rectification
power  is  arbitrarily large  for  sufficiently  small temperature  $\tau_{\rm
  min}$.

\section{Conclusions}
\label{sec:concl}

We  have discussed the  problem of  heat conduction  in classical  and quantum
low-dimensional  systems.   At   the  classical  level,  convincing  numerical
evidence exists  for the  validity of  the Fourier law  of heat  conduction in
linear mixing systems, {\em i.e.}, in systems without exponential instability.
As  a  consequence,  the  exponential  instability  (Lyapunov  chaos)  is  not
necessary for the establishment of  the Fourier law.  Moreover, breaking total
momentum conservation  is crucial for the  validity of the  Fourier law while,
somehow surprisingly, a  less important role seems to be  played by the degree
of dynamical chaos.

At  the quantum  level,  where the  motion  is characterized  by  the lack  of
exponential dynamical instability, we have  shown that the Fourier law of heat
conduction can be derived from  a pure quantum dynamical evolution without any
additional  statistical   assumptions.   Similarly  to   our  observations  in
classical models, our results for a chain of interacting spins suggest that in
quantum mechanics, which is characterized by the lack of exponential dynamical
instability, the  onset of quantum chaos  is required for the  validity of the
Fourier law.

We have  also discussed the  phenomenon of thermal rectification  in
different classical  models  and  discussed  different types  of
microscopic  classical mechanisms that  lead to rectification of
heat flow.  We have  focused on the magnetically-induced thermal
rectification and showed that for this mechanism, the rectification
ratio $\Delta$ is arbitrarily large for sufficiently small
temperature of one of the heat   baths.    Present days  nano-scale
experiments with meso-scopic devices should allow implementation of
our theoretical  model.

\begin{theacknowledgments}
  We  gratefully   acknowledge  support   by  the  MIUR-PRIN   2005  ``Quantum
  computation with  trapped particle arrays,  neutral and charged''. C.  M.-M.
  acknowledges  a  Lagrange  fellowship  from  the  Institute  for  Scientific
  Interchange Foundation.
\end{theacknowledgments}

\bibliographystyle{aipproc}

\end{document}